\documentclass{article}

\usepackage{arxiv}
\usepackage[utf8]{inputenc} 
\usepackage[T1]{fontenc}    
\usepackage{nicefrac}       
\usepackage{microtype}      
\usepackage{lipsum}		
\usepackage{doi}

\usepackage{graphicx} 
\usepackage{booktabs} 
\usepackage[font=large,labelfont=bf]{caption} 

\usepackage{amsfonts, amsmath, amsthm, amssymb} 
\usepackage{wrapfig} 
\usepackage{comment}
\usepackage{bm}
\usepackage{anyfontsize}
\usepackage{multirow}
\usepackage{tikz}
\usepackage{bbm}
\usepackage{url}
\PassOptionsToPackage{hyphens}{url}\usepackage{hyperref}
\usepackage{subfigure}
\usepackage{parskip}
\usepackage{float} 
\usepackage{natbib}
\usepackage[section]{placeins}
\usepackage{xurl}

\title{Spatial Correlation in Weather Forecast Accuracy: \\ A Functional Time Series Approach}

\author{Phillip A.~Jang\thanks{Cornell University, Center for Applied Mathematics, 657 Frank H.T. Rhodes Hall, Ithaca, NY 14853} \\
	Center for Applied Mathematics\\
	Cornell University\\
	Ithaca, NY 14853 \\
	\And
	David S.~Matteson\thanks{Cornell University, Department of Statistics and Data Science, 1196 Comstock Hall, Ithaca, NY 14853} \\
	Department of Statistics and Data Science\\
	Cornell University\\
	Ithaca, NY 14853 \\
}

\renewcommand{\shorttitle}{Spatial Correlation in Weather Forecast Accuracy: A Functional Time Series Approach}

\hypersetup{
pdftitle={DataExpoFunctionalTimeSeriesCOST},
pdfsubject={stat.ME},
pdfauthor={Phillip A.~Jang, David S.~Matteson},
pdfkeywords={Functional Data Analysis, Spatiotemporal, Random Effects, GARCH, Weather Forecast Data, Data Expo 2018},
}

\begin{document}
\maketitle

\begin{abstract}
A functional time series approach is proposed for investigating spatial correlation in daily maximum temperature forecast errors for 111 cities spread across the U.S. The modelling of spatial correlation is most fruitful for longer forecast horizons, and becomes less relevant as the forecast horizon shrinks towards zero. For 6-day-ahead forecasts, the functional approach uncovers interpretable regional spatial effects, and captures the higher variance observed in inland cities versus coastal cities, as well as the higher variance observed in mountain and midwest states. The functional approach also naturally handles missing data through modelling a continuum, and can be implemented efficiently by exploiting the sparsity induced by a B-spline basis.

The temporal dependence in the data is modeled through temporal dependence in functional basis coefficients. Independent first order autoregressions with generalized autoregressive conditional heteroskedasticity [AR(1)+GARCH(1,1)] and Student-t innovations work well to capture the persistence of basis coefficients over time and the seasonal heteroskedasticity reflecting higher variance in winter. Through exploiting autocorrelation in the basis coefficients, the functional time series approach also yields a method for improving weather forecasts and uncertainty quantification. The resulting method corrects for bias in the weather forecasts, while reducing the error variance.
\end{abstract}


\newpage
\section{Introduction}
\label{intro}
A \emph{functional time series} is a time-indexed sequence of stochastic processes $\{f_t(\tau)\}_{t=1}^\infty$ where each $f_t(\cdot)$ is a random function on the domain $\mathcal{T}$. By unifying functional data analysis with time series analysis, it presents an approach to modelling randomness on curves, surfaces, and other phenomena varying over a spatial continuum where these functional data are observed regularly over time and exhibit serial dependence. \cite{Ramsay05} and \cite{Tsay10} provide background for functional data analysis and time series analysis respectively, while \cite{Hormann12} provides background for functional time series. 

\cite{Hyndman07} and \cite{Hyndman08} propose a forecasting approach for functional time series and apply it to demographic data. \cite{Hyndman09} proposes a weighted functional approach which assigns more weight to recent observations and yields an improvement in forecast accuracy. Further developments in functional forecasting with applications to demographic data are found in \cite{Hyndman11}, \cite{Hyndman13}, and \cite{Dokumentov18}. \cite{Aue17} develops the theory for functional GARCH models. \cite{Kowal17} introduces a Bayesian framework for functional time series, and \cite{Kowal19} develops functional autoregression for sparsely and irregularly sampled data.

\begin{figure}
\centering
\includegraphics[width=\textwidth]{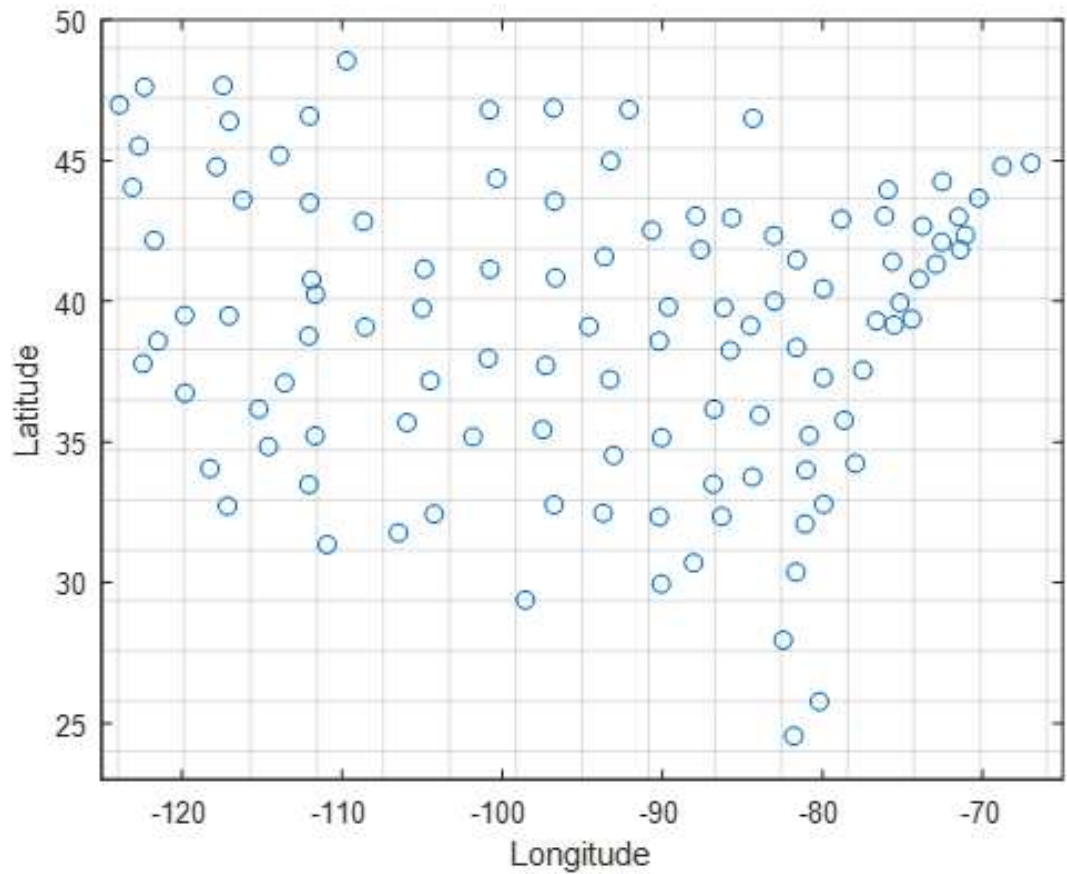}
\caption{Locations of cities included in weather forecast data shown in blue. The grey grid lines indicate the locations of the knots chosen for the B-spline bases.}
\label{fig:ForecastLocations}
\end{figure}

The most common examples of functional data are samples of 1-dimensional curves, and a contribution of this paper is the analysis of a model for 2-dimensional surface data. In this application, the spatial domain $\mathcal{T}$ is a rectangle in $\mathbb{R}^2$ containing the range of longitudes and latitudes covering the lower 48 states. The data set is taken from the 2018 American Statistical Association Data Expo\footnote{\url{http://ww2.amstat.org/sections/graphics/datasets/DataExpo2018.zip}}, consisting of daily maximum temperature forecasts from the National Weather Service for 111 cities spread across the US (excluding Alaska and Hawaii) over the period from July 2014 to September 2017. The locations of the cities are illustrated in Figure \ref{fig:ForecastLocations}. Forecasts range from same-day to six-days-ahead and are compared to actual temperature recorded at city airports.

A functional time series approach is applied to investigate and extract the structure of the spatial correlation in forecast errors. Through modelling the entire continuum instead of individual points, the proposed functional data approach also naturally handles missing data. This is a vital benefit, as data records are frequently incomplete and forecasts are not always available at every location.  This paper also extends the methodology of \cite{Hyndman07} with the incorporation of a heteroskedastic time series model to account for the higher unpredictability of weather in winters compared to summers.

Through exploiting both the spatial and temporal dependence in weather forecast errors, one can estimate the next day's expected forecast error and correct the forecast accordingly. 

In this paper, a four-step procedure is used to facilitate this estimation. Section \ref{splinesmooth} describes the initial pre-processing step, where we first construct continuous surface data from the discrete observations using manually selected B-splines. The resolution of the chosen B-splines
are set so that cities are able to be grouped into regions of interest. Once these surfaces are built, the second step is to reduce dimension to a smaller spatial basis which describes the most important modes of variation, and this is described in Section \ref{spatialbf}. After the spatial basis is chosen, the next step is to model the temporal dependence in the random coefficients which scale the basis functions, and this is described in Section \ref{randcoef}. The time series model facilitates the final step, the prediction of forecast errors, and this is described in Section \ref{predicterror}.

\section{Methodology}

We first fix a forecast horizon $h$ of interest, in this case between 0 to 6 days ahead. For the given forecast horizon, let $Y_t(\tau)$ be the forecast error of maximum temperature on day $t$ for the city located at spatial coordinates $\tau \equiv (longitude,latitude)$. Spatial correlation is captured through the following spatio-temporal random effect model:
\begin{equation}
Y_t(\tau) = \mu(\tau) + \sum_{k=1}^K \beta_{kt}\varphi_k(\tau) + \varepsilon_t(\tau).
\end{equation}
Here, $\mu(\tau)$ represents the mean forecast error and is assumed fixed over time. Estimation of the mean function is described in Section \ref{splinesmooth}. The spatial basis functions $\varphi_1(\tau), \ldots, \varphi_K(\tau)$ describe the main modes of variation in the forecast errors, capturing spatial dependence across different regions of the U.S. The construction of the basis functions is described in Sections \ref{splinesmooth} and \ref{spatialbf}. The random coefficients $\beta_{kt}$ capture temporal correlation. They are modelled using independent AR(1)+GARCH(1,1) processes, as described in Section \ref{randcoef}. Lastly, $\varepsilon_t(\tau)$ is a white noise process independent of the random coefficients, and is assumed i.i.d. $N(0,\sigma^2)$ for all $t$ and $\tau$. It is used to capture, including any measurement errors, the remaining variation not explained by the spatio-temporal random effect.

Writing $\Phi(\tau) = [\varphi_1(\tau) \cdots \varphi_K(\tau)]^T$ and $\bm{\beta}_t = [\beta_{1t} \cdots \beta_{Kt}]^T$, this model implies the following spatio-temporal covariance function:
\begin{equation}
Cov(Y_t(\tau),Y_{t'}(\tau')) = \Phi(\tau)^T \mathbb{E}[\bm{\beta}_t\bm{\beta}_{t'}^T] \Phi(\tau') + \sigma^2 \mathbbm{1}_{\{t=t',\tau=\tau'\}},
\end{equation}
where the indicator function $\mathbbm{1}_{\{t=t',\tau=\tau'\}}$ is $1$ when both $t=t'$ and $\tau=\tau'$, and $0$ otherwise.

\subsection{Pre-Processing with Spline Smoothing\label{splinesmooth}}

A two-step procedure is employed to construct the spatial basis functions $\varphi_k$. As an initial pre-processing step, we first need to construct continuous surface data from the discretely observed values using manually selected B-splines, where the knot sequence is at the right level of tightness to capture regional groupings of the cities. Once these surfaces are built, the singular value decomposition (SVD) is used in Section \ref{spatialbf} to reduce dimension and identify the $\varphi_k$ which describe the most important modes of variation.

For simplicity, we consider the spatial domain as a subset of $\mathbb{R}^2$. Specifically, we define a 2-D cubic B-spline basis over the rectangle $[-124,-66]\times[24,49]$, which contains the range of longitudes and latitudes covering the lower 48 states. The 2-D splines are built from the tensor product of 1-D cubic B-splines on longitude and latitude individually. Refer to \cite{deBoor01} for further background on B-splines. The MATLAB package `bspline' by \cite{Hunyadibsp} is used to implement the splines.

For this particular dataset, knot sequences with 13 equally spaced interior knots were able to capture interesting regional groupings of the cities, and the knots are visualized against the cities by the grey grid lines in Figure \ref{fig:ForecastLocations}. This results in 17 cubic B-splines in each dimension, and 289 2-D splines in the resulting tensor product, denoted by $S_1(\tau),\ldots,S_{289}(\tau)$, with 

\begin{align*}
\mathcal{K}^\text{Lon} &= [-124,-124,-124,-124,-119.86,-115.71,\ldots,-66,-66,-66,-66]\text{,}\\
\mathcal{K}^\text{Lat} &= [24,24,24,24,25.79,27.57,\ldots,49,49,49,49]\text{, and}
\end{align*}
\begin{equation*}
\begin{bmatrix}S_1(lon,lat)\\ \vdots \\ S_{289}(lon,lat)\end{bmatrix} = \begin{bmatrix}B_{1,\mathcal{K}^\text{Lon}}(lon)\\ \vdots \\ B_{17,\mathcal{K}^\text{Lon}}(lon)\end{bmatrix} \otimes \begin{bmatrix}B_{1,\mathcal{K}^\text{Lat}}(lat)\\ \vdots \\ B_{17,\mathcal{K}^\text{Lat}}(lat)\end{bmatrix}.
\end{equation*}

For each day $t=1,\ldots,m$, we fit splines such that the coefficients $\hat{c}_{t,1},\ldots,\hat{c}_{t,289}$ solve the following optimization problem:
\begin{equation}
\min_{c_{t,1},\ldots,c_{t,289}} \sum_{i=1}^{n_t} \left[Y_t(\tau_{t,i}) - \sum_{j=1}^{289} c_{t,j}S_j(\tau_{t,i})\right]^2\text{,}
\end{equation}
where $\tau_{t,1},\ldots,\tau_{t,n_t}$ are the observation locations available on day $t$. This handles (moderate amounts of) missing data naturally since missing observations $Y_t(\tau)$ are simply omitted from the objective function. Also, since B-splines have compact support, the resulting system is sparse and can be solved efficiently. 

As the number of coefficients exceeds the number of observation locations, a truncated singular value decomposition is used to solve for the coefficients. The truncated SVD acts as a form of regularization similar to ridge regression, and further details can be found in \cite{hansen1987}. In Figure \ref{fig:ForecastLocations}, there is a lack of observations in regions of the rectangle outside the borders of the United States. The truncated SVD yields stable results by imposing zero coefficients for the splines that lie in these data-free regions.

We denote the resulting coefficient matrix by $C = [\hat{c}_{t,j}]$. We use the column means $\bar{C} = \begin{bmatrix}\bar{c}_1,\ldots,\bar{c}_{289}\end{bmatrix}$ to estimate the mean function $\mu(\tau)$ as:
\begin{equation}
\hat{\mu}(\tau) = \bar{\bar{Y}} + \sum_{j=1}^{289} \bar{c}_j S_j(\tau)
\end{equation}

where $\bar{\bar{Y}} = \left(\sum_{t=1}^m n_t\right)^{-1}\sum_{t=1}^m\sum_{i=1}^{n_t} Y_t(\tau_{t,i})$ is the sample mean of all observed forecast errors.

\subsection{Constructing Spatial Basis Functions $\varphi_k$\label{spatialbf}}

As each spline function captures variation specific to a sub-rectangle of the domain, the amount of information in each region of the domain is encoded through the coefficients of its corresponding splines. Through studying the covariance of the B-spline coefficients, we are able to piece together variation across multiple sub-rectangles to study the important regional effects in the weather forecast errors. A principal component analysis yields a ranked order of groupings through the eigendecomposition of the covariance matrix, or equivalently, the singular value decomposition of the mean-centered data matrix. The eigenvectors of the covariance matrix identify weightings of the spline coefficients which contribute the most variation, with said variation quantified by the size of the eigenvalue.

After mean-centering the columns of $C$, the singular value decomposition $C-\bar{C} = U\Sigma V^T$ provides principal component loadings as the columns of $V$, allowing for dimension reduction that is mean-square optimal in the coefficient space $\text{Col}(C-\bar{C})$. Refer to \cite{jolliffe2002principal} for further background on principal component analysis and singular value decomposition. Assuming the SVD is written in descending order of singular values, the spatial basis functions $\varphi_k(\tau)$ are built using the first $K$ columns of $V$, representing the $K$ most important principal components:
\begin{equation}
\varphi_k(\tau) = \sum_{j=1}^{289}V_{jk}S_j(\tau) \text{ for } k = 1,\ldots,K \ll 289.
\end{equation}

\begin{figure}
\centering
\includegraphics[width=0.85\textwidth]{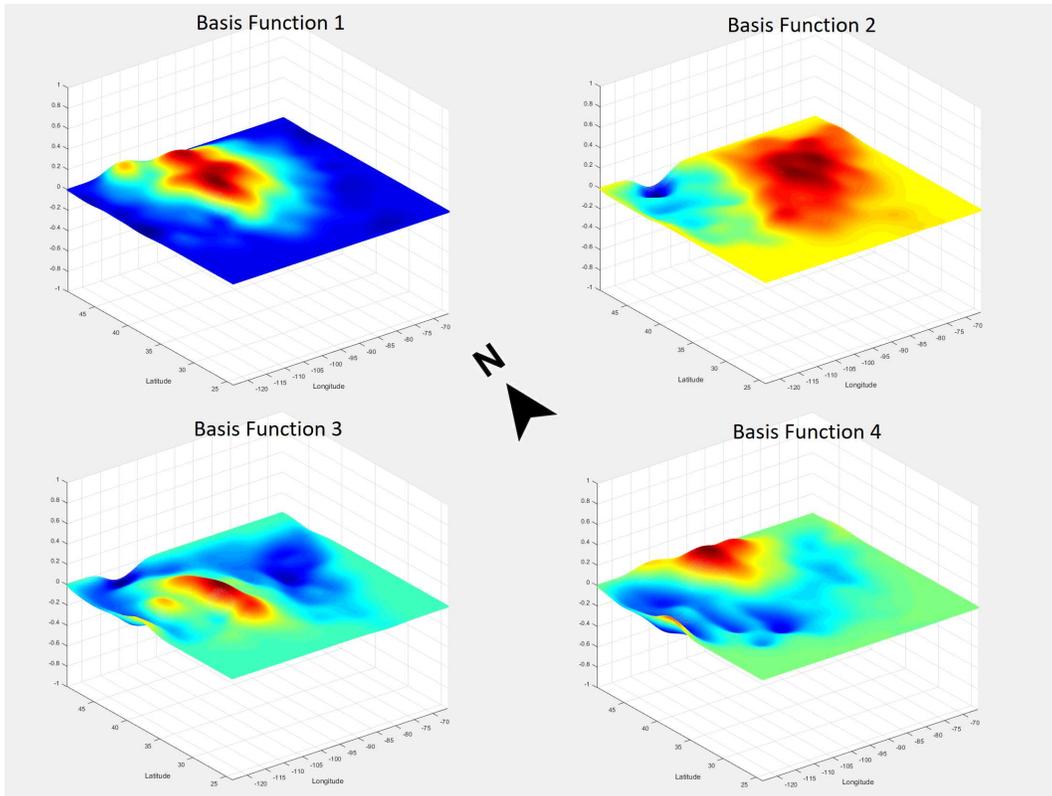}
\caption{\vspace{-0.2cm}First 4 basis functions for 6-day-ahead forecasts.}
\label{fig:BF6Day}
\end{figure}

The first four basis functions are shown in Figure \ref{fig:BF6Day} for the 6-day-ahead forecast horizon. The first basis function represents an inland versus coastal effect, as cities further inland had greater variance in their forecast errors compared to coastal cities. The second basis function represents an east versus west effect, with the opposing signs of the regions allowing for a differentiation between the regions. The third and fourth basis functions represent mountain state and midwest state effects respectively, as these regions have the most unpredictable weather.

\begin{figure}
\centering
\includegraphics[width=0.85\textwidth]{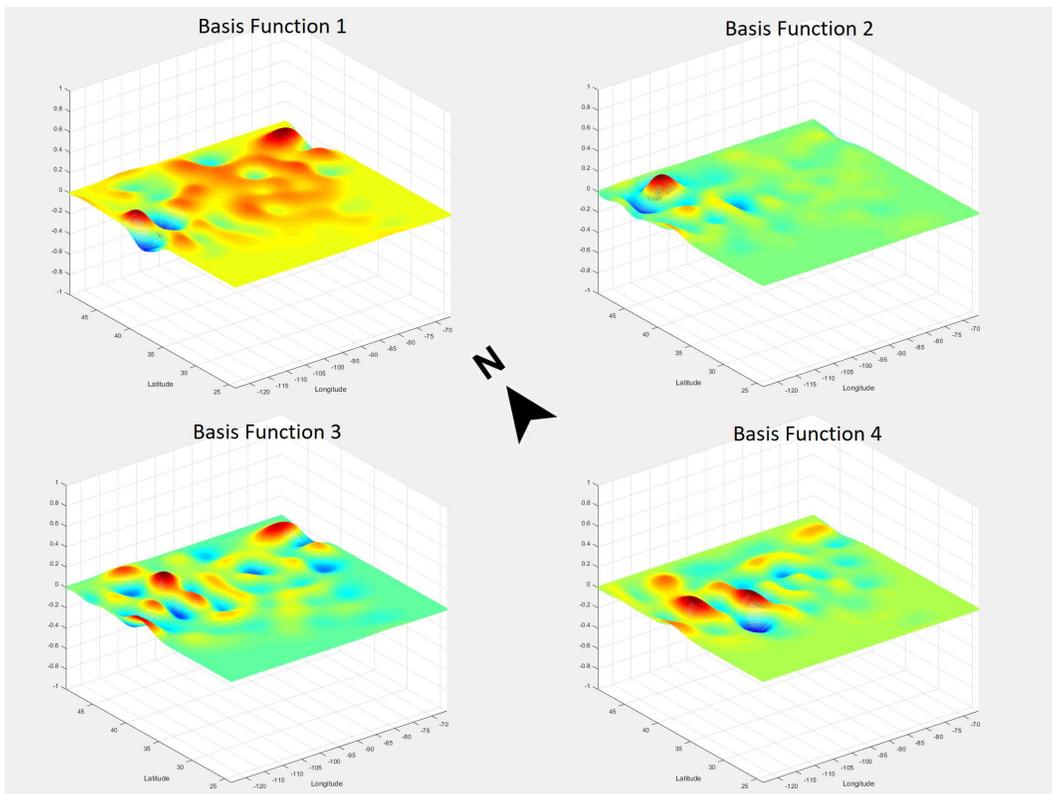}
\caption{\vspace{-0.2cm}First 4 basis functions for same-day (0-day-ahead) forecasts.}
\label{fig:BFSameDay}
\end{figure}

This structure of spatial correlation is most prevalent in 6-day-ahead forecasts, but vanishes as the forecast horizon shrinks to zero. For example, Figure \ref{fig:BFSameDay} shows the first four basis functions for same-day forecasts, and these basis functions lack any coherent spatial structure.

For the 6-day-ahead forecasts, a cutoff of $K=20$ basis functions is used. Each of the 20 selected basis functions ranged from 9\% to 1.4\% of explained variance, while the basis functions after the 20th each contribute less than 1.4\% of explained variance and revealed no interesting spatial structure. The results of Table \ref{tab:errorcomp} were stable above $K=20$ basis functions, justified by Figure \ref{fig:BFSelection}.

\FloatBarrier

\FloatBarrier

\clearpage
\subsection{Modelling Random Coefficients\label{randcoef}}

Once the spatial basis functions are defined as above, coefficients $\beta_{1t},\ldots,\beta_{Kt}$ for each day $t$ are estimated on this reduced $K$-dimensional basis as solutions to the following optimization problem:
\begin{equation}
\min_{\beta_{1t},\ldots,\beta_{Kt}} \sum_{i=1}^{n_t} \left[Y_t(\tau_{t,i}) - \hat{\mu}(\tau_{t,i}) - \sum_{k=1}^K \beta_{kt}\varphi_k(\tau_{t,i})\right]^2.
\end{equation}

Temporal dependence in forecast errors is then modelled through these coefficients. Empirically, an AR(1)+GARCH(1,1) model with Student-t innovations was found to provide a good description of the observed coefficients. Specifically,

\begin{align}
\beta_{kt} &= \psi_k\beta_{k,t-1} + u_{kt}, \ & u_{kt}|\mathcal{F}_{t-1} \sim t_{\nu_k}(0,\eta_{kt}^2)\\
\eta_{kt}^2 &= \omega_k + \alpha_k u_{k,t-1}^2 + \gamma_k \eta_{k,t-1}^2\
\end{align}

where $\mathcal{F}_{t-1}$ is the information set up to time $t-1$ (the $\sigma$-field generated by $u_{k,t-1}, u_{k,t-2}, \ldots$ for all $k$). The $u_{1t}, \ldots, u_{Kt}$ are assumed conditionally independent given $\mathcal{F}_{t-1}$. The proposed model implies the following conditional mean and covariance functions:

\begin{align}
\mathbb{E}[Y_t(\tau)|\mathcal{F}_{t-1}] &= \mu(\tau) + \sum_{k=1}^K \psi_k\beta_{k,t-1}\varphi_k(\tau)\\
Cov(Y_t(\tau),Y_t(\tau')|\mathcal{F}_{t-1}) &= \sum_{k=1}^K \frac{\nu_k}{\nu_k-2}\left[\omega_k + \alpha_k u_{k,t-1}^2 + \gamma_k \eta_{k,t-1}^2\right]\varphi_k(\tau)\varphi_k(\tau') + \sigma^2 \mathbbm{1}_{\{\tau=\tau'\}}.
\end{align}

\begin{figure}
\centering
\includegraphics[width=\textwidth]{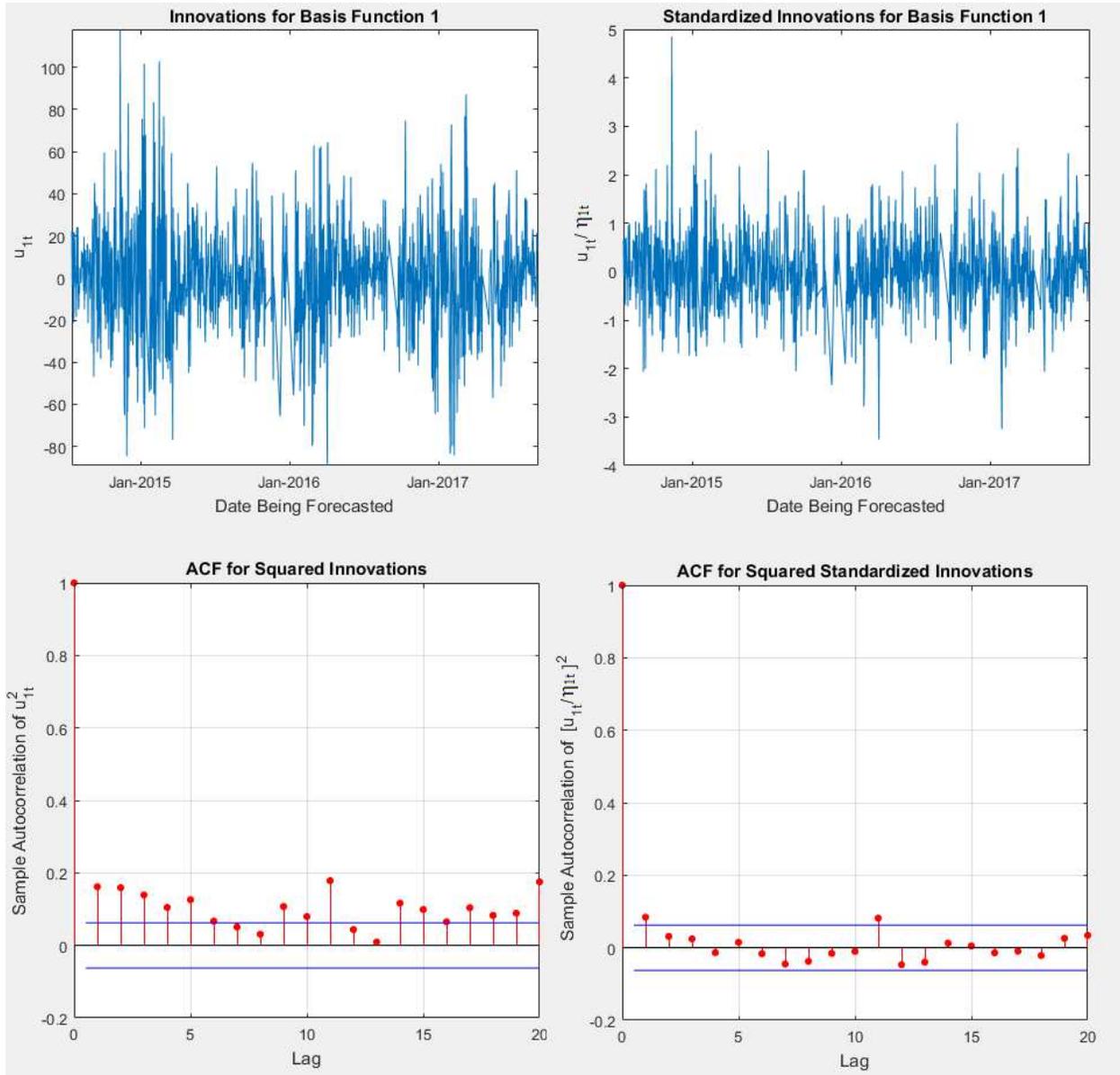}
\caption{Estimated innovations $u_{1t}$ (top left) and standardized innovations $u_{1t}/\eta_{1t}$ (top right) for basis function 1 at forecast horizon 6-days-ahead, and their respective sample autocorrelation functions (bottom, left and right, respectively). Approximate 95\% pointwise confidence intervals are drawn as horizontal lines in the autocorrelation plots.}
\label{fig:GARCHDiagnostic}
\end{figure}

The resulting innovations for the first basis function $u_{1t}$ are shown in Figure \ref{fig:GARCHDiagnostic} on the top left. Notably, the innovations exhibit a seasonal heteroskedasticity with winter weather being the most unpredictable. The GARCH process characterizes the heteroskedasticity well, as the standardized innovations $u_{1t}/\eta_{1t}$ exhibit approximately constant variance, and the squared standardized innovations show no significant autocorrelation.

For the 6-day-ahead weather forecasts, the AR+GARCH parameter estimates for the first four basis functions are shown in Table \ref{paramest}. All fitted models were stationary with similar amounts of autocorrelation based on the similar values of $\psi_k$. Furthermore, all exhibit a high persistence in variance, indicated by the large values of $\gamma_k$. The earlier basis functions had conditional distributions with heavier tails, indicated by lower values of $\nu_k$.


\begin{table}[H]
\centering
\caption{AR(1)+GARCH(1,1) parameter estimates for first four basis functions for 6-day-ahead forecasts.}\label{paramest}
\begin{tabular}{lrrrrr}
\hline
\multicolumn{1}{|l|}{Basis Function 1}   & \multicolumn{1}{c|}{$\psi_1$}          & \multicolumn{1}{c|}{$\omega_1$} & \multicolumn{1}{c|}{$\alpha_1$}        & \multicolumn{1}{c|}{$\gamma_1$}        & \multicolumn{1}{c|}{$\nu_1$}           \\ \hline
\multicolumn{1}{|l|}{Estimate}           & \multicolumn{1}{r|}{0.65}              & \multicolumn{1}{r|}{13.50}      & \multicolumn{1}{r|}{0.09}              & \multicolumn{1}{r|}{0.89}              & \multicolumn{1}{r|}{8.33}              \\ \hline
\multicolumn{1}{|l|}{Approx. Std. Error} & \multicolumn{1}{r|}{0.02}              & \multicolumn{1}{r|}{6.91}       & \multicolumn{1}{r|}{0.02}              & \multicolumn{1}{r|}{0.02}              & \multicolumn{1}{r|}{2.05}              \\ \hline
\multicolumn{1}{|l|}{t-ratio}            & \multicolumn{1}{r|}{26.91}             & \multicolumn{1}{r|}{1.95}       & \multicolumn{1}{r|}{4.25}              & \multicolumn{1}{r|}{37.14}             & \multicolumn{1}{r|}{4.06}              \\ \hline
\multicolumn{1}{|l|}{p-value}            & \multicolumn{1}{r|}{\textless{}0.0001} & \multicolumn{1}{r|}{0.0507}     & \multicolumn{1}{r|}{\textless{}0.0001} & \multicolumn{1}{r|}{\textless{}0.0001} & \multicolumn{1}{r|}{\textless{}0.0001} \\ \hline
                                         & \multicolumn{1}{l}{}                   & \multicolumn{1}{l}{}            & \multicolumn{1}{l}{}                   & \multicolumn{1}{l}{}                   & \multicolumn{1}{l}{}                   \\ \hline
\multicolumn{1}{|l|}{Basis Function 2}   & \multicolumn{1}{c|}{$\psi_2$}          & \multicolumn{1}{c|}{$\omega_2$} & \multicolumn{1}{c|}{$\alpha_2$}        & \multicolumn{1}{c|}{$\gamma_2$}        & \multicolumn{1}{c|}{$\nu_2$}           \\ \hline
\multicolumn{1}{|l|}{Estimate}           & \multicolumn{1}{r|}{0.57}              & \multicolumn{1}{r|}{18.67}      & \multicolumn{1}{r|}{0.05}              & \multicolumn{1}{r|}{0.92}              & \multicolumn{1}{r|}{10.90}             \\ \hline
\multicolumn{1}{|l|}{Approx. Std. Error} & \multicolumn{1}{r|}{0.03}              & \multicolumn{1}{r|}{11.43}      & \multicolumn{1}{r|}{0.02}              & \multicolumn{1}{r|}{0.03}              & \multicolumn{1}{r|}{3.95}              \\ \hline
\multicolumn{1}{|l|}{t-ratio}            & \multicolumn{1}{r|}{21.08}             & \multicolumn{1}{r|}{1.63}       & \multicolumn{1}{r|}{3.00}              & \multicolumn{1}{r|}{34.18}             & \multicolumn{1}{r|}{2.76}              \\ \hline
\multicolumn{1}{|l|}{p-value}            & \multicolumn{1}{r|}{\textless{}0.0001} & \multicolumn{1}{r|}{0.1024}     & \multicolumn{1}{r|}{0.0027}            & \multicolumn{1}{r|}{\textless{}0.0001} & \multicolumn{1}{r|}{0.0057}            \\ \hline
                                         & \multicolumn{1}{l}{}                   & \multicolumn{1}{l}{}            & \multicolumn{1}{l}{}                   & \multicolumn{1}{l}{}                   & \multicolumn{1}{l}{}                   \\ \hline
\multicolumn{1}{|l|}{Basis Function 3}   & \multicolumn{1}{c|}{$\psi_3$}          & \multicolumn{1}{c|}{$\omega_3$} & \multicolumn{1}{c|}{$\alpha_3$}        & \multicolumn{1}{c|}{$\gamma_3$}        & \multicolumn{1}{c|}{$\nu_3$}           \\ \hline
\multicolumn{1}{|l|}{Estimate}           & \multicolumn{1}{r|}{0.53}              & \multicolumn{1}{r|}{7.11}       & \multicolumn{1}{r|}{0.03}              & \multicolumn{1}{r|}{0.96}              & \multicolumn{1}{r|}{13.31}             \\ \hline
\multicolumn{1}{|l|}{Approx. Std. Error} & \multicolumn{1}{r|}{0.03}              & \multicolumn{1}{r|}{6.46}       & \multicolumn{1}{r|}{0.01}              & \multicolumn{1}{r|}{0.02}              & \multicolumn{1}{r|}{4.87}              \\ \hline
\multicolumn{1}{|l|}{t-ratio}            & \multicolumn{1}{r|}{19.95}             & \multicolumn{1}{r|}{1.10}       & \multicolumn{1}{r|}{2.32}              & \multicolumn{1}{r|}{47.14}             & \multicolumn{1}{r|}{2.73}              \\ \hline
\multicolumn{1}{|l|}{p-value}            & \multicolumn{1}{r|}{\textless{}0.0001} & \multicolumn{1}{r|}{0.2709}     & \multicolumn{1}{r|}{0.0203}            & \multicolumn{1}{r|}{\textless{}0.0001} & \multicolumn{1}{r|}{0.0063}            \\ \hline
                                         & \multicolumn{1}{l}{}                   & \multicolumn{1}{l}{}            & \multicolumn{1}{l}{}                   & \multicolumn{1}{l}{}                   & \multicolumn{1}{l}{}                   \\ \hline
\multicolumn{1}{|l|}{Basis Function 4}   & \multicolumn{1}{c|}{$\psi_4$}          & \multicolumn{1}{c|}{$\omega_4$} & \multicolumn{1}{c|}{$\alpha_4$}        & \multicolumn{1}{c|}{$\gamma_4$}        & \multicolumn{1}{c|}{$\nu_4$}           \\ \hline
\multicolumn{1}{|l|}{Estimate}           & \multicolumn{1}{r|}{0.57}              & \multicolumn{1}{r|}{13.01}      & \multicolumn{1}{r|}{0.06}              & \multicolumn{1}{r|}{0.91}              & \multicolumn{1}{r|}{14.75}             \\ \hline
\multicolumn{1}{|l|}{Approx. Std. Error} & \multicolumn{1}{r|}{0.03}              & \multicolumn{1}{r|}{7.56}       & \multicolumn{1}{r|}{0.02}              & \multicolumn{1}{r|}{0.03}              & \multicolumn{1}{r|}{6.97}              \\ \hline
\multicolumn{1}{|l|}{t-ratio}            & \multicolumn{1}{r|}{21.21}             & \multicolumn{1}{r|}{1.72}       & \multicolumn{1}{r|}{3.50}              & \multicolumn{1}{r|}{31.04}             & \multicolumn{1}{r|}{2.11}              \\ \hline
\multicolumn{1}{|l|}{p-value}            & \multicolumn{1}{r|}{\textless{}0.0001} & \multicolumn{1}{r|}{0.0853}     & \multicolumn{1}{r|}{0.0005}            & \multicolumn{1}{r|}{\textless{}0.0001} & \multicolumn{1}{r|}{0.0344}            \\ \hline
\end{tabular}
\end{table}

\newpage
\section{Empirical Performance}

\begin{figure}[!h]
\centering
\includegraphics[width=\textwidth]{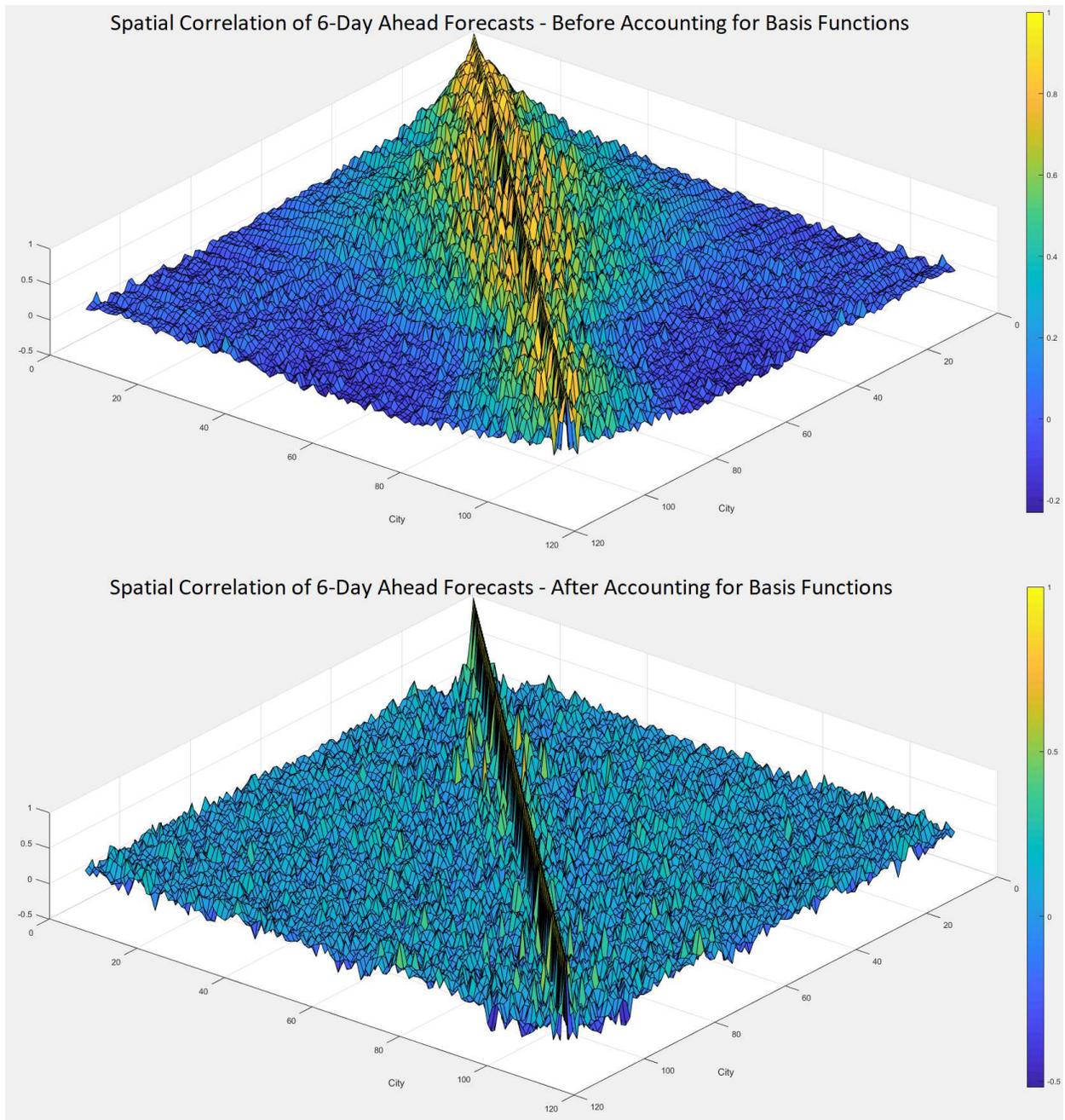}
\caption{Spatial correlation in 6-day-ahead forecast errors for before (top) and after (bottom) accounting for basis functions.}
\label{fig:SpaCorrBeforeAfter}
\end{figure}

For each pair of cities $(i,j)$, sample spatial correlation for the 6-day-ahead forecasts is computed before and after accounting for the spatial basis functions. More specifically, given a pair of cities located at $\tau_i$ and $\tau_j$, the top of Figure \ref{fig:SpaCorrBeforeAfter} shows
\begin{equation}
\rho_{i,j}^{\text{before}}=\frac{\sum_{t}[Y_t(\tau_i)-\bar{Y}_\cdot(\tau_i)][Y_t(\tau_j)-\bar{Y}_\cdot(\tau_j)]}{\sqrt{\sum_{t}[Y_t(\tau_i)-\bar{Y}_\cdot(\tau_i)]^2 \sum_{t}[Y_t(\tau_j)-\bar{Y}_\cdot(\tau_j)]^2}},
\end{equation}

and the bottom of Figure \ref{fig:SpaCorrBeforeAfter} shows 
\begin{equation}
\rho_{i,j}^{\text{after}}=\frac{\sum_{t}[\hat{\varepsilon}_t(\tau_i)-\bar{\hat{\varepsilon}}_\cdot(\tau_i)][\hat{\varepsilon}_t(\tau_j)-\bar{\hat{\varepsilon}}_\cdot(\tau_j)]}{\sqrt{\sum_{t}[\hat{\varepsilon}_t(\tau_i)-\bar{\hat{\varepsilon}}_\cdot(\tau_i)]^2 \sum_{t}[\hat{\varepsilon}_t(\tau_j)-\bar{\hat{\varepsilon}}_\cdot(\tau_j)]^2}},
\end{equation}

for the residuals $\hat{\varepsilon}_t(\tau) = Y_t(\tau) - \hat{\mu}(\tau) - \sum_{k=1}^K \beta_{kt}\varphi_k(\tau)$. Above, the sums are over the days $t$ with no missing observations, and $\bar{Y}_\cdot(\tau_i)$ and $\bar{\hat{\varepsilon}}_\cdot(\tau_i)$ are sample averages of $Y_t(\tau_i)$ and $\hat{\varepsilon}_t(\tau)$ respectively over such $t$.

The cities are numbered from 1 to 111 (shown on the $x$- and $y$-axes), and cities are ordered from east to west, resulting in a concentration of high correlation along the main diagonal in the first figure. After accounting for $K=20$ spatial basis functions, the second figure indicates the lack of spatial correlation in the residuals and provides evidence that the proposed model provides an adequate approximation of the observed spatial correlation in forecast errors. 

\begin{figure}[!h]
\centering
\includegraphics[width=\textwidth]{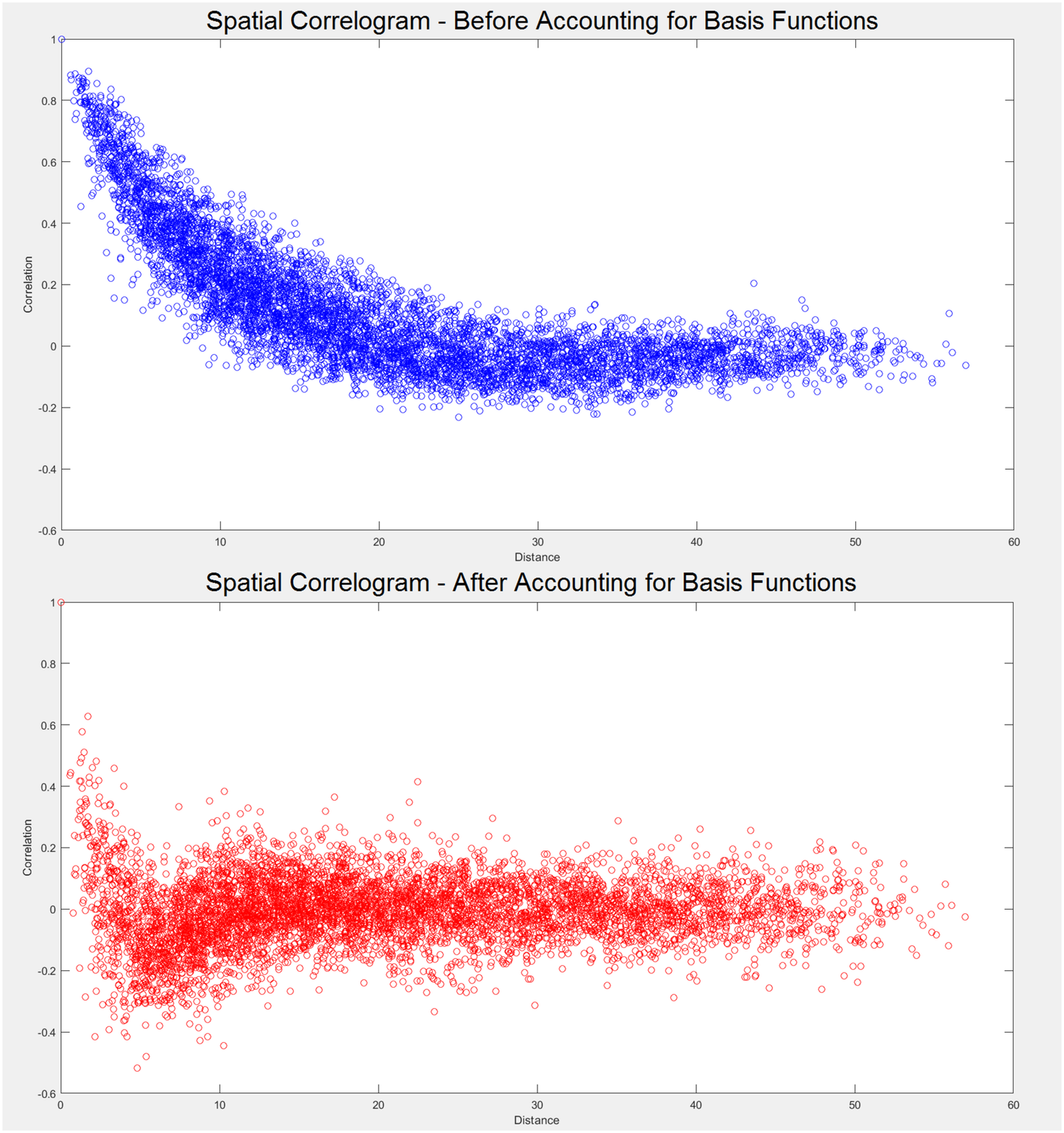}
\caption{Correlogram of 6-day-ahead forecast errors for before (top) and after (bottom) accounting for basis functions.}
\label{fig:CorreloBeforeAfter}
\end{figure}

\begin{figure}[!h]
    \centering
    \includegraphics[width=\textwidth]{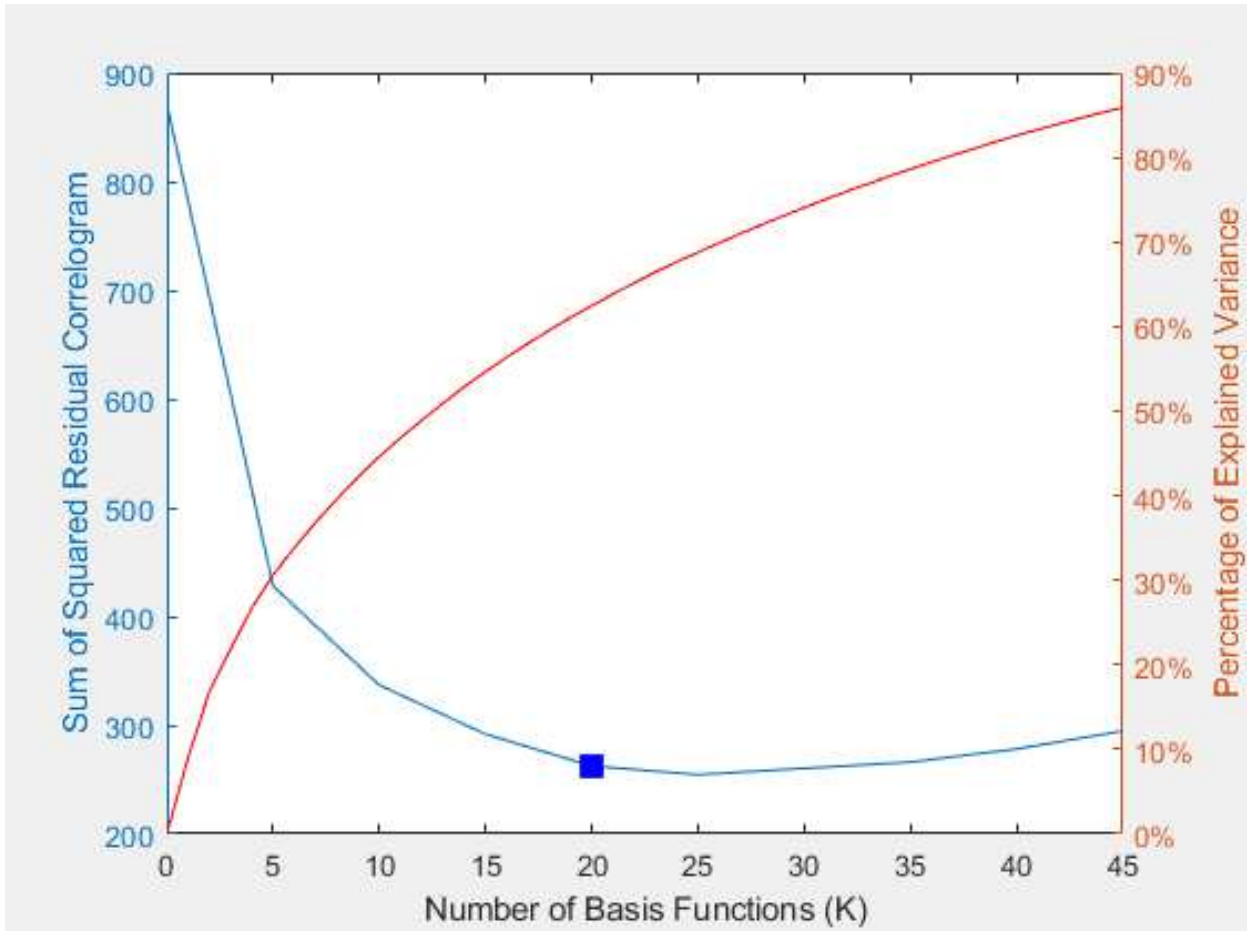}
    \caption{Sum of squared residual correlogram $\sum_{i,j}(\rho_{i,j}^{\text{after}})^2$ (blue) for a changing number of basis functions $K$. After $K=20$ basis functions, the Frobenius norm of the residual spatial correlation matrix stops decreasing, indicating the remaining basis functions capture local variation rather than regional variation.}
    \label{fig:BFSelection}
\end{figure}

The correlograms before and after accounting for the basis functions are shown in Figure \ref{fig:CorreloBeforeAfter}, visualizing the sample spatial correlations against city distance. The cutoff of $K=20$ is justified in Figure \ref{fig:BFSelection}, where the squared Frobenius norm of the residual spatial correlation matrix ($\sum_{i,j}(\rho_{i,j}^{\text{after}})^2$) is computed for different values of $K$. The Frobenius norm stops significantly decreasing after $K=20$ indicating that additional basis functions do not explain any more spatial correlation, and capture individual city variation rather than regional variation.

\FloatBarrier
\subsection{Predicting Forecast Errors to Improve Forecast Accuracy\label{predicterror}}

Through exploiting autocorrelation in the coefficients, the AR+GARCH parameter estimates can be used to predict the next day's basis coefficients, as
\begin{align}
\beta_{k,t+1}|\mathcal{F}_{t} &\sim t_{\nu_k}(\psi_k \beta_{kt},\eta_{k,t+1}^2)\text{, and}\\
\eta_{k,t+1}^2 &= \omega_k + \alpha_k u_{kt}^2 + \gamma_k \eta_{kt}^2\text{.}
\end{align}

Substituting the parameter estimates $\hat{\psi}_k$, $\hat{\omega}_k$, $\hat{\alpha}_k$, $\hat{\gamma}_k$, and $\hat{\nu}_k$ in place of the true parameters yields an approximate distribution which can be used for prediction and uncertainty quantification.

By setting the predicted coefficient to $\hat{\beta}_{k,t+1} = \hat{\psi}_k \beta_{kt}$, this can then be used to predict next day's weather forecast errors by setting
\begin{equation}
\hat{Y}_{t+1}(\tau) = \hat{\mu}(\tau) + \sum_{k=1}^K \hat{\beta}_{k,t+1}\varphi_k(\tau).
\end{equation}

These predicted errors $\hat{Y}_{t+1}(\tau)$ can then be used to adjust the next day's weather forecast (of the same horizon). More explicitly, for forecast horizon $h$ and location $\tau$, if $F_{t,t+h}(\tau)$ is the forecast on day $t$ of the maximum temperature on day $t+h$ and $A_{t+h}(\tau)$ is the actual temperature on day $t+h$ so that $Y_t(\tau) = F_{t,t+h}(\tau)-A_{t+h}(\tau)$, we can define an adjusted forecast for day $t+1$ as
\begin{equation}
F_{t+1,t+1+h}^{adj}(\tau) = F_{t+1,t+1+h}(\tau)-\hat{Y}_{t+1}(\tau).
\end{equation}

The improvement on weather forecasts can then be assessed by comparing the adjusted errors
\begin{equation}
Z_{t+1}(\tau) = F_{t+1,t+1+h}^{adj}(\tau)-A_{t+1+h}(\tau)
\end{equation}
to the unadjusted errors $Y_{t+1}(\tau)$.

The mean and variance of $Y$ and $Z$ across all $t$ and $\tau$ is shown in Table \ref{tab:errorcomp}.

\begin{table}
\centering
\caption{Comparison of 6-day ahead weather forecast errors before and after adjustment using $K=20$ basis functions.}
\label{tab:errorcomp}
\begin{tabular}{lrr}
\hline\noalign{\smallskip}
 & $Y_t(\tau)$ & $Z_t(\tau)$ \\
\noalign{\smallskip}\hline\noalign{\smallskip}
Mean & -1.17 & 0.00 \\
Standard Deviation & 6.61 & 5.82 \\
\noalign{\smallskip}\hline
\end{tabular}
\end{table}

\begin{figure}
\centering
\includegraphics[width=\textwidth]{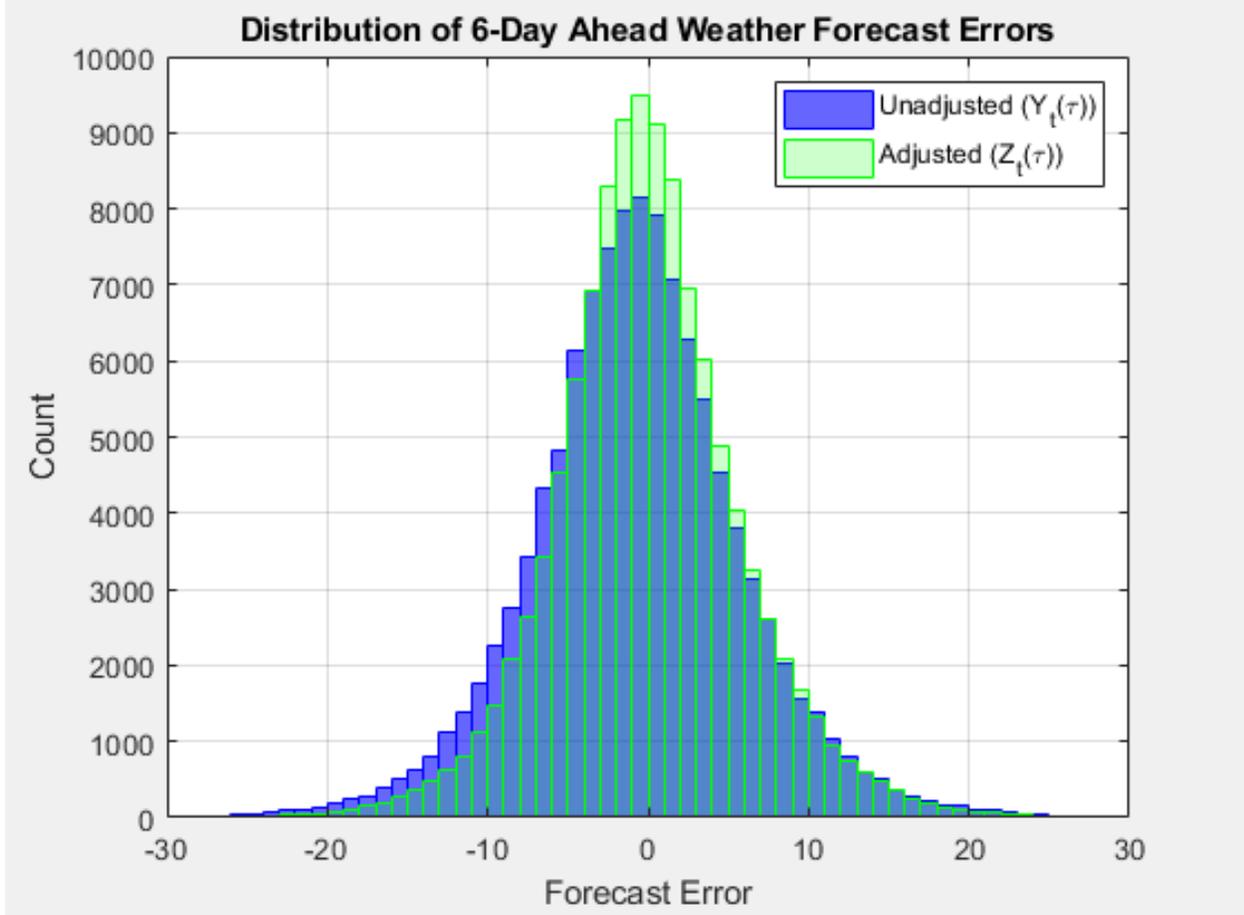}
\caption{Comparison of distributions of 6-day ahead weather forecast errors before and after adjustment using $K=20$ basis functions.}
\label{fig:ErrorDist}
\end{figure}

The change in mean error indicates that the proposed method eliminates bias in the weather forecasts, while the reduction of standard deviation by $12\%$ indicates a significant average improvement in forecast accuracy. The shift in error distribution is illustrated in Figure \ref{fig:ErrorDist}, with the adjusted distribution centered at 0 and with smaller variance.

\section{Code}
For reproducing our analysis and figures, MATLAB code is available at:

\url{https://github.com/pjang23/dataexpo2018-functionaltimeseries}

To accomodate those without the MATLAB Econometrics Toolbox\texttrademark, the MATLAB package `ARMAX-GARCH-K-SK' by \citet{GabrielsenGARCH} is included as an alternative for fitting the AR+GARCH models.

\section{Conclusions}

We have introduced a functional time series approach to investigating spatial correlation in weather forecast accuracy. The modelling of spatial correlation is most fruitful for the longer forecast horizons, and becomes less relevant as the forecast horizon shrinks towards zero. For 6-day-ahead weather forecasts, the functional approach uncovers interpretable regional spatial effects, and captures the higher variance observed in inland cities versus coastal cities, as well as the higher variance observed in mountain and midwest states. The functional approach also naturally handles missing data and can be implemented efficiently by exploiting the sparsity induced by using a B-spline basis.

Independent first order autoregressions with generalized autoregressive conditional heteroskedasticity [AR(1)+GARCH(1,1)] and Student-t innovations worked well to capture the persistence of coefficients over time and the seasonal heteroskedasticity reflecting higher variance in winter. Autocorrelation in the basis coefficients can further be exploited to improve weather forecasts, especially at longer horizons, and the resulting approach eliminates bias while reducing error variance.

\section{Acknowledgements}
This work is supported in part by the Natural Sciences and Engineering Research Council of Canada (PGS-D 502888), the National Science Foundation (1455172, 1934985, 1940124, 1940276, and 2114143), a Xerox PARC Faculty Research Award, the United States Agency for International Development (USAID), and Cornell University Atkinson Center for a Sustainable Future (AVF-2017).

\bibliographystyle{spbasic}
\bibliography{DataExpoFunctionalTimeSeriesCOST}

\end{document}